\documentclass[conference]{IEEEtran}
\IEEEoverridecommandlockouts
\usepackage{comment}
\usepackage{cite}
\usepackage{amsmath,amssymb,amsfonts}
\usepackage{algorithmic}
\usepackage{graphicx}
\usepackage{textcomp}
\usepackage{xcolor}
\def\BibTeX{{\rm B\kern-.05em{\sc i\kern-.025em b}\kern-.08em
    T\kern-.1667em\lower.7ex\hbox{E}\kern-.125emX}}
\begin{document}

\title{A Deep Learning Architecture with Spatio-Temporal Focusing for Detecting Respiratory Anomalies
}
\author{Dat~Ngo$^{1}$, 
             Lam~Pham$^{2}$, 
             Huy~Phan*$^{3}$, 
             Minh~Tran$^{4}$,
             Delaram~Jarchi$^{1}$
             \\ \\
1. School of Computer Science and Electronic Engineering, University of Essex, UK.\\
\{dn22678, delaram.jarchi\}@essex.ac.uk \\            
2. Center for Digital Safety \& Security, Austrian Institute of Technology, Austria. \{lam.pham@ait.ac.at\} \\

3. Amazon Alexa, Cambridge, MA, US. \{huypq@amazon.co.uk\}  \\

4. Nuffield Department of Clinical Neurosciences, University of Oxford, UK. \{minh.tran@ndcn.ox.ac.uk\}\\

 \thanks{(*) The work was done when Huy Phan was at the School of Electronic Engineering and Computer Science, Queen Mary University of London, UK, and prior to joining Amazon Alexa.}
}

\maketitle

\begin{abstract}
This paper presents a deep learning system applied for detecting anomalies from respiratory sound recordings. Our system initially performs audio feature extraction using Continuous Wavelet transformation. This transformation converts the respiratory sound input into a two-dimensional spectrogram where both spectral and temporal features are presented. Then, our proposed deep learning architecture inspired by the Inception-residual-based backbone performs the spatio-temporal-focusing and multi-head attention mechanism to classify respiratory anomalies. In this work, we evaluate our proposed models on the benchmark SPRSound (The Open-Source SJTU Paediatric Respiratory Sound) database proposed by the IEEE BioCAS 2023 challenge. As regards the Score computed by an average between the average score and harmonic score, our robust system has achieved Top-1 performance with Scores of 0.810, 0.667, 0.744, and 0.608 in Tasks 1-1, 1-2, 2-1, and 2-2, respectively.
\end{abstract}


\section{Introduction}
\label{intro}

Many respiratory diseases such as tuberculosis, asthma, chronic obstructive pulmonary disease (COPD), and lower respiratory tract infection (LRTI), have resulted in a significant and concerning mortality rate of 6.2 million individuals worldwide~\cite{who_1}. In the UK, chronic respiratory illnesses cause the hospitalization of over 700,000 people annually~\cite{lunguk}. During the initial phase, respiratory diseases cause damage or obstruction to the airways in the lungs, resulting in a restricted flow of air during both the inhalation and exhalation processes. Therefore, machine learning systems (i.e. including conventional machine learning models and deep learning architectures) are proposed for analyzing anomalies in respiratory sounds caused by damage or obstruction in the lung's airways. These systems can facilitate clinicians to diagnose these diseases at an earlier stage in the most scalable, noninvasive, and time-saving workflow.
In general, there are two main groups of machine learning systems used for classifying respiratory anomalies. As the first group makes use of handcrafted features, a variety of techniques such as statistical features~\cite{palaniappan2013machine}, entropy-based features~\cite{zhang2009novel}, Mel Frequency Cepstral Coefficients (MFCCs)~\cite{jung2021efficiently} are exploited to transform lung sounds into feature vectors. Next, conventional machine learning models explore these vectors to detect anomalies in lung sounds. Otherwise, the second group transforms the audio recordings into two-dimensional spectrograms. These spectrograms such as S-transform~\cite{moukadem2013robust}, MFCC spectrogram~\cite{datta2017automated}, and log-mel spectrogram~\cite{garcia2020detecting, 9909611} are generated to capture spectral and temporal information of respiratory sounds. Next, these spectrograms are inputted into network architectures such as convolutional neural network (CNN) based architectures~\cite{pham2021cnn, ngo2021deep} or recurrent neural network (RNN) based architectures~\cite{ic_cnn_19_iccas, perna2019deep} for classification. While MFCC and log-mel spectrograms were found as a popular representation, the fixed window size is still hindering a proper resolution for feature extraction. To overcome this, an alternative way of using Wavelet-based spectrogram~\cite{akay1997wavelet, 9871440} with a better multi-resolution analysis is proposed thanks to its suitability in adjusting both temporal window length and the wide frequency range across the length.

In this paper, we leverage our previous work of Inception-residual-based architecture~\cite{pham2022wider} to classify anomalies from respiratory sounds. However, we propose spatio-temporal-focusing and multi-head attention mechanisms to explore the effect of spatio-temporal information across different temporal lengths of spectrograms. To demonstrate our robust performance in detecting respiratory anomalies, we evaluate our proposed systems on the IEEE BioCAS 2023 challenge. Our contributions are as follows: (1) We investigated multiple spectrograms extracted from Continuous Wavelet transform with three different mother waves of Amor, Morse, and Bump at different temporal dimensions. (2) We successfully applied Inception-residual-based architecture combined with spatio-temporal-focusing and multi-head attention mechanisms to explore various spectrograms at different temporal lengths and pinpoint that the more features explored on temporal information, the more efficiency in detecting anomalies from respiratory sounds.

\section{SPRSound database and tasks definition}
\label{dataset}
\subsection{SPRSound database}


In this work, we utilize the 2022 SPRSound: Open-Source SJTU Paediatric Respiratory Sound database, which was collected in the Shanghai Children’s Medical Center (SCMC), China~\cite{zhang2022sprsound}. The database comprises 2,683 audio recordings obtained from 292 patients aged between 1 month and 18 years old. These audio recordings were recorded by Yunting model II Stethoscope at 8 KHz sampling rate with 16-bit precision. Each recording underwent a careful examination by experts, who labeled each recording as Poor Quality (PQ), Normal (N), Continuous Adventitious Sound (CAS), Discontinuous Adventitious Sound (DAS), or CAS and DAS (CD). In addition, respiratory experts also annotated the onset (starting time) and offset (ending time) of every audio event within the recordings. As a result, this database  consists of audio events classified as Normal (N), Rhonchi (Rho), Wheeze (W), Stridor (Str), Wheeze and Crackle (B), Coarse Crackle (CC), and Fine Crackle (FC). Furthermore, these audio recordings and events show various duration ranging from 0.304\,s-15.36\,s and 0.126\,s to 7.152\,s, respectively. The imbalance in the distribution among classes in both recording and event levels also makes the classification more challenging.

\subsection{Tasks Definition and Evaluating Setup}

There are two levels (i.e. event and recording) of classification tasks in SPRSound database, referred to as Task 1 and Task 2. \textbf{Task 1} comprises two sub-tasks of Task 1-1 and Task 1-2. While Task 1-1 is to classify the respiratory sound events as Normal and Adventitious, Task 1-2 involves classifying these events into \textit{N}, \textit{Rho}, \textit{W}, \textit{Str}, \textit{CC}, \textit{FC}, or \textit{B}. 
\textbf{Task 2} focuses on the entire recording, which is also separated into Task 2-1 and Task 2-2. 
In particular, Task 2-1 aims to classify the respiratory recordings as Normal, Adventitious, and Poor Quality. Meanwhile, Task 2-2 is a multi-class classification, where the respiratory recordings are classified into \textit{N}, \textit{CAS}, \textit{DAS}, \textit{CD}, or \textit{PQ}. We adhere to the evaluation metrics as mentioned in~\cite{zhang2022sprsound}, every task and its sub-tasks in this paper are evaluated by \textit{sensitivity} (SE), \textit{specificity} (SP), \textit{average score} (AS), \textit{harmonic score} (HS), and the average of AS and HS (Score). As the SPRSound database is proposed in the IEEE BioCAS 2023 challenge, we follow the challenge to separate the database into Training and Validation sets. After evaluating our proposed models on the Validation set, we submit the models to the challenge for evaluation on a blind Test set.








\section{The Proposed System}
\label{proposedframework}

\begin{figure}[h]
	\centering
	\centerline{\includegraphics[width=0.75\linewidth]{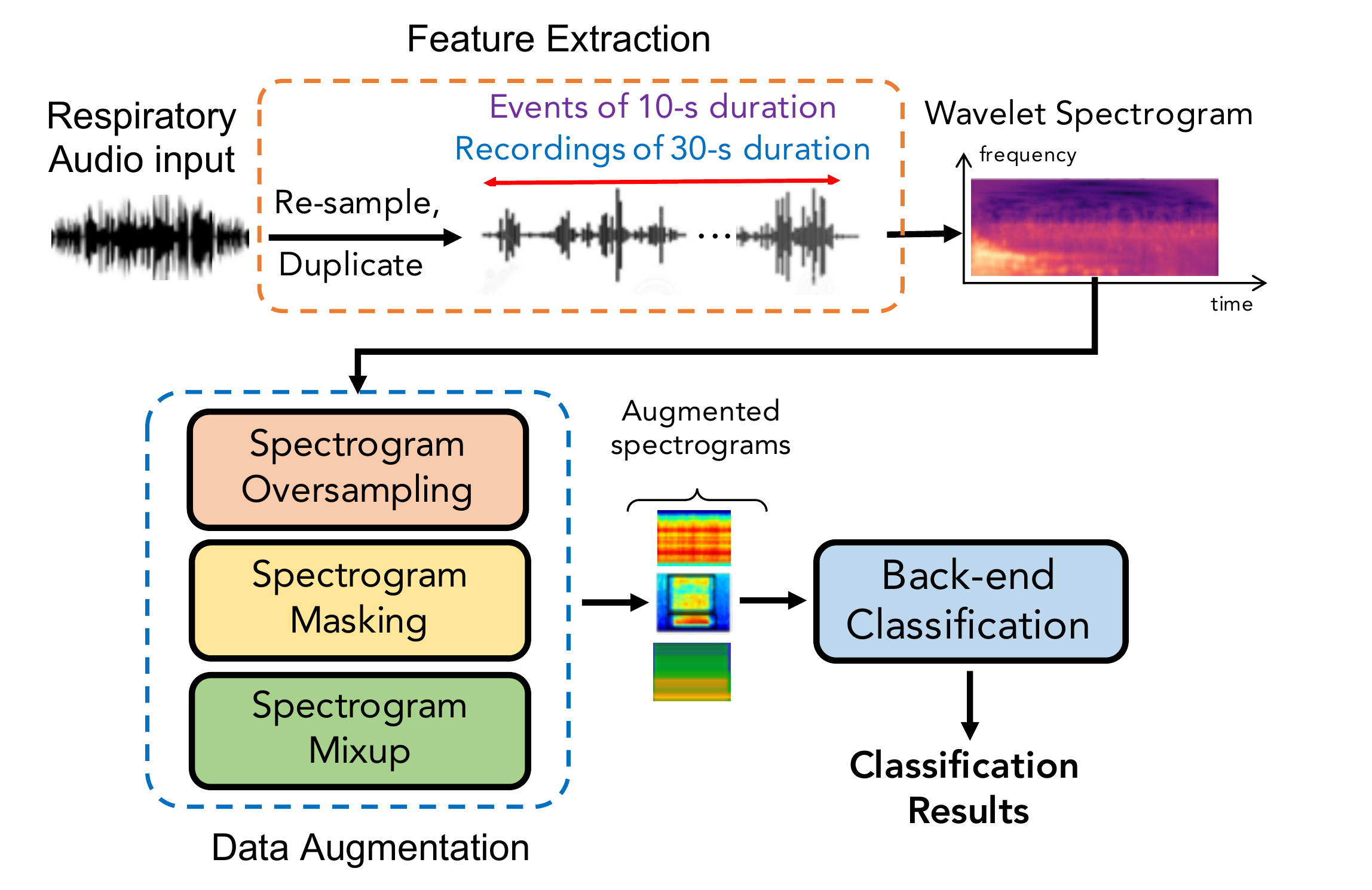}}
	 \vspace{-0.3 cm}
	\caption{The high-level architecture of our proposed system for detecting anomalies in respiratory audio inputs.}
	\vspace{-0.4 cm}
	\label{fig:overall_framework}
\end{figure}

Overall, our proposed system for detecting anomalies in respiratory sounds as shown in Fig.~\ref{fig:overall_framework} consists of three main steps: low-level spectrogram feature extraction, data augmentation, and back-end classification.

\subsection{The low-level spectrogram feature extraction}

\begin{figure*}[t]
	\centering
	\centerline{\includegraphics[width=0.7\linewidth]{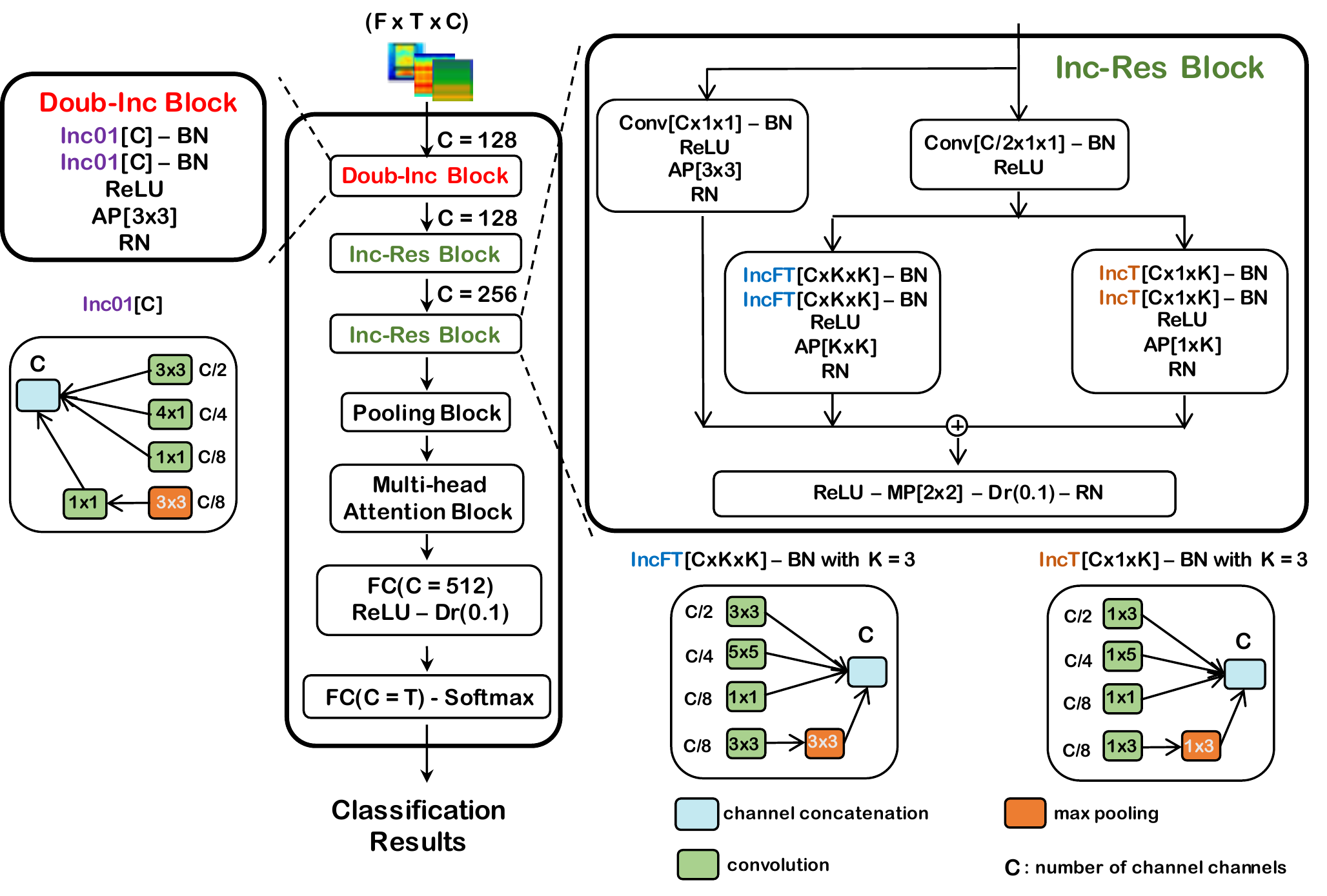}}
	 \vspace{-0.3 cm}
	\caption{The proposed architecture for the Back-end classification }
	\vspace{-0.2 cm}
	\label{fig:backbone_archi}
\end{figure*}	

The respiratory events/recordings are resampled to 4 kHz as abnormal respiratory sounds typically have frequency bands within the range of 60-2000 Hz~\cite{zhang2022sprsound}. Next, while the resampled respiratory events with different lengths are duplicated to ensure they have a consistent length of 10 seconds, audio recordings are also duplicated to make sure that all audio recordings have a consistent duration of 30 seconds. A band-pass filter of 60-2000 Hz is then applied to suppress background noise in each respiratory event/recording. Finally, these respiratory events/recordings undergo a transformation process to generate two-dimensional spectrograms. This transformation applies Continuous Wavelet transformation, which employs Amor, Bump, and Morse as the Wavelet mother functions. As a result of these transformations, three types of spectrograms are obtained from each respiratory event/recording. Finally, while all spectrograms of the event level are scaled to different sizes of $128{\times}128$, $128{\times}256$, and $128{\times}512$, each type of spectrogram of the recording level is scaled to the different sizes of $140{\times}256$, $140{\times}512$, $140{\times}1024$ (i.e. frequency bands${\times}$the number of time frames).

\subsection{Data Augmentation}

To address the issue of imbalanced data discussed in Section~\ref{dataset}, we propose three data augmentation techniques on both the event and entire recording levels after spectrograms are generated from the feature extraction step. First, we randomly oversample the spectrograms to ensure an equal number of spectrograms per class in each batch size. This helps to balance the representation of different classes in the training data. Second, spectrograms within each batch are then randomly cropped with a reduction of 10 bins in both the time and frequency dimensions. This encourages the learning process to focus on the partial loss of information at each dimension~\cite{park2019specaugment}. Third, we apply the mixup data augmentation technique~\cite{tokozume2017learning} to increase the variation in the training data. This technique combines pairs of spectrograms from different classes, creating new synthetic samples. The goal is to enhance the diversity of the training data and enlarge Fisher's criterion (i.e. the ratio of between-class distance to within-class variance in the feature space). Eventually, the augmented spectrograms are fed into a back-end classifier, reporting the classification results.

\subsection{The back-end classification}

The proposed architecture for the back-end classification is illustrated in Fig.~\ref{fig:backbone_archi}, which is based on Inception-residual-based network architecture inspired by our prior work in~\cite{pham2022wider}. 
In detail, our proposed network architecture comprises one Doub-Inc Block, two Inc-Res Blocks, one Pooling Block, one Multi-head Attention Block, and two fully-connected layers. 
The Doub-Inc Block includes two Inc01 blocks followed by Batch Normalization (BN), Rectified Linear Unit (ReLU), Average Pooling (AP[Kernel Size]), Dropout (Dr(Drop Ratio), and Residual Normalization (RN( $\lambda$ = 0.4)) inspired from\cite{kim2022qti}). 
The Inc01 block is constructed as a variant of the naive inception layer~\cite{szegedy2015going} with fixed kernel sizes of [3×3], [1×1], and [4×1]. 
Next, two Inc-Res Blocks share the same architecture as inspired by our prior work in~\cite{pham2022wider}, but channel numbers increase from 128 to 256 to form a deeper view of the channel dimension. 
The detailed structure for every Inc-Res Block is described in the right-hand side of Fig.~\ref{fig:backbone_archi}, which presents two main blocks of IncFT[Channel×Kernel Size] block and IncT[Channel×Kernel Size] block, and layers of Conv, BN, Dr, ReLU, AP, Max Pooling (MP[Kernel Size]), and RN( $\lambda$ = 0.4). 
Notably, IncFT[Channel×Kernel Size] and IncT[Channel×Kernel Size] layers propose different kernel sizes defined as [K×K] and [1×K]. The value of K is changed as details on the bottom part of Fig.~\ref{fig:backbone_archi}. 
The idea of IncFT[Channel×Kernel Size] is proposed to strengthen the network to learn effectively not only the widespread frequency bands but also the distribution of energy in certain frequency bands across the temporal length of the input spectrograms. 
Additionally, we suppose that anomalies in lung sounds come from different duration with different changes in their frequency, which might cause more changes in the temporal dimension of each spectrogram rather than in certain frequency bands. 
Therefore, the IncT[Channel×Kernel Size] is proposed to focus completely on temporal information from different dimensions of the time frame in spectrogram inputs. Following each IncFT[Channel×Kernel Size] and IncT[Channel×Kernel Size] block, a ReLU, an AP layer, and RN layer are applied before adding sub-branch results together. 
The Pooling Block makes use of global pooling layers to extract three features from the second Inc-Res Block: (1) global average pooling across the channel dimension, (2) global max pooling across the temporal dimension, and (3) global average pooling across frequency dimensions. 

Furthermore, it has been observed in~\cite{vaswani2017attention} that individual attention heads acquire distinct sets of weight matrices. When these self-attention heads are combined, they create a multi-head self-attention layer. This layer enhances the learning performance to generate more comprehensive embedding features among different anomalies in respiratory sounds. As a result, the output of the Pooling Block is then presented to the Multi-head Attention block as shown in Fig.~\ref{fig:backbone_archi}.  At each Attention block, we apply three multi-head attention layers on three dimensions of frequency, time, and channel. Each multi-head attention layer is configured to have 16 as the number of heads and 32 as the key dimension.
The output of each multi-head attention layer is a one-dimensional embedding feature. We then concatenate these features before feeding them into fully connected blocks. While the first dense layer comprises a fully connected layer (FC[C = $512$]) followed by a ReLU, Dr, the second dense layer comprises a fully connected layer (FC[C = $T$]) followed by a Softmax, where $T$ is defined according to the number of target classes.

\textbf{Model implementation and Training Loss functions:} We implement our proposed models in this paper using the Tensorflow framework. As we propose to use mixup data augmentation, the labels are not one-hot format.
Therefore, we use Kullback-Leibler (KL) divergence loss in the proposed networks as shown in Eq. (\ref{eq:kl_loss}) below:

\begin{align}
\label{eq:kl_loss}
Loss_{KL}(\Theta) = \sum_{n=1}^{N}\mathbf{y}_{n}\log(\frac{\mathbf{y}_{n}}{\mathbf{\hat{y}}_{n}})  +  \frac{\lambda}{2}||\Theta||_{2}^{2},
\end{align}

where \(Loss_{KL}(\Theta)\) is KL-loss function, $\Theta$ describes the trainable parameters of the network, $\lambda$ denotes the $\ell_2$-norm regularization coefficient experimentally set to 0.0001, \(N\) is the batch size, $\mathbf{y_{n}}$ and $\mathbf{\hat{y}_{n}}$  are the ground truth and the network output, respectively.

\section{The Experimental Results and Discussion}
\begin{table}[t]
    \caption{Performance comparison among spectrograms on the Validation set in Task 1-1 and Task 1-2 (event level)} 
        	\vspace{-0.2cm}
    \centering
    \scalebox{0.75}{
    \begin{tabular}{|c| c c c|c c c|c c|} 
        \hline 
             
             \textbf{System}&\multicolumn{3}{|c|}{\textbf{Task 1-1}} &\multicolumn{3}{|c|}{\textbf{Task 1-2}}\\
    
        \hline 
             &SE/SP  &AS/HS &Score  &SE/SP  &AS/HS &Score    \\
             \hline
             Wavelet (Amor)  &  &  & & & &  \\
             128$\times$128  &0.73/0.86  &0.79/0.79   &0.79      &0.57/0.86  &0.72/0.68 &0.70    \\
             128$\times$256  &0.77/0.87  &0.82/0.82   &0.82      &0.61/0.87  &0.74/0.72 &0.73    \\
             128$\times$512  &0.76/0.90  &0.83/0.82   &0.83      &0.64/0.90  &0.77/0.75 &0.76    \\
             \hline
             Wavelet (Bump)  &  &  & & & &  \\
             128$\times$128  &0.77/0.90  &0.83/0.83   &0.83     &0.64/0.90  &0.77/0.75 &0.76    \\
             128$\times$256  &0.79/0.88  &0.83/0.83   &0.83     &0.67/0.88  &0.78/0.76 &0.77    \\
             128$\times$512  &0.81/0.91  &0.86/0.86   &\textbf{0.86}     &0.67/0.92  &0.79/0.78 &\textbf{0.79}    \\
             \hline
             Wavelet (Morse) &  &  & & & &  \\
             128$\times$128  &0.80/0.84  &0.82/0.82   &0.82     &0.63/0.84  &0.74/0.72 &0.73    \\
             128$\times$256  &0.75/0.92  &0.83/0.83   &0.83     &0.62/0.92  &0.77/0.74 &0.76    \\
             128$\times$512  &0.84/0.87  &0.85/0.85   &0.85     &0.69/0.87  &0.78/0.77 &0.77    \\
            
       \hline 
    \end{tabular}
                       }
        \vspace{-0.1cm}
    \label{table:event} 
\end{table}
\textbf{On Event Level (Task 1-1 and Task 1-2 on Validation set):} As the experimental results are shown in Table~\ref{table:event}, it can be seen that when the number of time frames in each type of Wavelet spectrogram is extended (i.e. from 128 to 256 and 512), the performance is further improved in both Task 1-1 and Task 1-2. 
This can be explained as when more temporal information is provided, the proposed model has the greater ability to exploit effectively information across the temporal length of each spectrogram.
For instance, extending from 128 to 256 and 512 helps Wavelet (Bump) increase from 0.83 to 0.86 in Score in Task 1-1 and achieves an improvement of 3\% in Score in Task 1-2.  
Compare among three evaluating spectrograms, Wavelet (Bump) outperforms Wavelet (Morse) and Wavelet (Amor), and achieves the highest Scores at the size of 128$\times$512 in both Task 1-1 and Task 1-2 with 0.86 and 0.79, respectively. 
It indicates that Wavelet (Bump) spectrogram is the most appropriate representation to capture distinct features of respiratory events.

\textbf{On Record Level (Task 2-1 and Task 2-2 on Validation set):} Similar to Task 1-1 and Task 1-2, the increase of time frames (ranging from 256 to 512 and 1024) in each type of Wavelet spectrogram results in improvements in both Task 2-1 and Task 2-2 as shown in Table~\ref{table:recording}. 
This again indicates the benefits of widening time frames in every spectrogram and the efficiency of our proposed model which is developed to focus on temporal information. 
For instance, the significant improvement of 15\% in AS and 13\% in HS are obtained in Task 2-1 when Wavelet (Bump) is extended from 140$\times$256 to 140$\times$1024. 
Notably, when the spectrograms are set to 140$\times$1024, Wavelet (Morse) outperforms Wavelet (Amor) and Wavelet (Bump),  and achieves the highest performance of Score in both Task 2-1 (at 0.71) and Task 2-2 (at 0.55).
This indicates that Wavelet (Morse) spectrogram is the most suitable representation for respiratory recordings. 

\textbf{On IEEE BioCAS 2023 grand challenge (All tasks):} Given the experimental results on both the event level and the entire encoding level, we indicate that our models with Wavelet (Bump) at the size 128$\times$512 and Wavelet (Morse) with the size of 140$\times$1024 achieve the best Scores on Task 1 and Task 2, respectively.
We, therefore, use these systems to submit to the IEEE BioCAS 2023 grand challenge and evaluate our systems with the blind Test set~\cite{scoreboard}.
As the results shown in Table~\ref{table:scoreboard}, our proposed model trained on Wavelet (Bump) at size 128$\times$512 has achieved Top-1 performance in tasks of event level. 
In particular, we gain the highest Score of 0.810 and 0.667 on Task 1-1 and Task 1-2, respectively. 
As regards the recording level, when the proposed model is trained with Wavelet (Morse) spectrogram at a size of 140$\times$1024, we surpass other systems to have the best Score of 0.608 in Task 2-2. 
Even though Task 2-1 has a lower Score of 0.744, compared to that in Top-2, our proposed system still outperforms other systems in general. 

\begin{table}[t]
    \caption{Performance comparison among spectrograms on Validation set in Task 2-1 and Task 2-2 (entire recording level)} 
        	\vspace{-0.2cm}
    \centering
    \scalebox{0.75}{
    \begin{tabular}{|c| c c c|c c c|c c|} 
        \hline 
             
             \textbf{System}&\multicolumn{3}{|c|}{\textbf{Task 2-1}} &\multicolumn{3}{|c|}{\textbf{Task 2-2}}\\
    
        \hline 
             &SE/SP  &AS/HS &Score  &SE/SP  &AS/HS &Score    \\
             \hline
             Wavelet (Amor)  &  &  & & & &  \\
             140$\times$256    &0.66/0.59  &0.63/0.62   &0.62    &0.33/0.59  &0.46/0.43 &0.45    \\
             140$\times$512    &0.56/0.78  &0.67/0.65   &0.66    &0.35/0.78  &0.56/0.48 &0.52    \\
             140$\times$1024   &0.77/0.64  &0.70/0.70   &0.70   &0.41/0.64  &0.52/0.50 &0.51   \\
             \hline
             Wavelet (Bump)  &  &  & & & &  \\
             140$\times$256   &0.54/0.57  &0.55/0.55   &0.55    &0.33/0.78  &0.55/0.46 &0.51    \\
             140$\times$512   &0.54/0.83  &0.69/0.65   &0.67    &0.32/0.83  &0.58/0.46 &0.52    \\
             140$\times$1024  &0.58/0.83  &0.70/0.68   &0.69   &0.32/0.85  &0.59/0.46 &0.53    \\
             \hline
             Wavelet (Morse) &  &  & & & &  \\
             140$\times$256   &0.60/0.76  &0.68/0.67   &0.67    &0.27/0.76  &0.52/0.40 &0.46    \\
             140$\times$512   &0.57/0.76  &0.66/0.65   &0.66    &0.35/0.76  &0.56/0.48 &0.52    \\
             140$\times$1024  &0.67/0.76  &0.71/0.71   &\textbf{0.71}   &0.40/0.76  &0.52/0.52 &\textbf{0.55}    \\
            
       \hline 
    \end{tabular}
                       }
    \label{table:recording} 
\end{table}

\begin{table}[t]
	\caption{Compare our proposed systems to the others submitted IEEE BioCAS 2023 challenge (Score(\%) on the blind Test set)}
    \centering
    \scalebox{0.80}{
    \begin{tabular}{|c| c c c c|} 
        \hline 
	   \textbf{Systems}    &\textbf{Task 1-1}  &\textbf{Task 1-2}   &\textbf{Task 2-1}  &\textbf{Task 2-2} \\
        \hline
	    \textbf{Top 1(Our system) }        &\textbf{0.810}             &\textbf{0.667}            &0.744       &\textbf{0.608}  \\
        \hline
            Top 2                     &0.733             &0.646            &\textbf{0.759}       &0.538     \\
	    Top 3                     &0.769             &0.632            &0.661       &0.512     \\
            Top 4                     &0.720             &0.593            &0.665       &0.549     \\
            Top 5                     &0.668             &0.555           &0.723       &0.524     \\
            Top 6                     &0.748             &0.599           &0.699       &0.411     \\
            Top 7                     &0.785             &0.648           &0.547       &0.417     \\
            Top 8                     &0.756             &0.467           &0.658       &0.458     \\
          \hline
    
    \end{tabular}
    }
    \vspace{-0.4cm}
    \label{table:scoreboard} 
\end{table}


\section{Conclusion}
\label{conclusion}
We have presented an Inception-residual-based network architecture supported by spatio-temporal-focusing and multi-head attention mechanisms for detecting respiratory anomalies. 
The results on IEEE BioCAS 2023 challenge, which  achieved the Top-1 performance, have proven that the efficiency of using higher time frames in the Wavelet (Bump) spectrogram for tasks on the event level and in the Wavelet (Morse) spectrogram for tasks on the recording level. 
In addition, the proposed model shows its ability in learning temporal information when the time frame of each spectrogram is extended. 





\end{document}